\documentclass[11pt,a4paper]{article}
\usepackage[english]{babel}
\usepackage[T1]{fontenc}
\usepackage[utf8]{inputenc}
\usepackage{graphicx}
\usepackage{mathtools, amssymb,hyperref,cite,doi}
\usepackage[hmargin=3cm,vmargin=3cm,bmargin=3cm]{geometry}
\begin{document}

\title{Magnetogenesis in Cyclical Universe}
\author{Natacha Leite$^\dagger$, Petar Pavlovi\'c$^\ddagger$}
\date{}

\maketitle
\center{ $^\dagger$Center for Space Research, North-West University, Mafikeng 2735, South Africa\\ natacha.leite@nwu.ac.za\\
 $^\ddagger$II. Institute for Theoretical Physics, Hamburg University, Luruper Chaussee 149, 22761 Hamburg, Germany\\  petar.pavlovic@desy.de}

\abstract{We propose a simple solution to the magnetogenesis problem based on cyclic cosmology. It is demonstrated that magnetic fields, of 
sufficient strengths to account for the present observational bounds, can be created
in the contracting phase preceding the beginning of the current cosmological cycle. The basic assumption of this model is that the Universe enters a contraction phase essentially empty, characterized
by small seed electric fields. In this framework, there is no need for any new theoretical additions to explain magnetogenesis, such as new scalar fields or the non-minimal 
coupling between curvature, scalar fields and the electromagnetic sector. Moreover, the proposed model is general in the sense that it does not assume any specific modified gravity 
theory to enable the cosmological bounce. When compared to the inflationary magnetogenesis paradigm, the proposed model also has the advantage that it does not lead to the backreaction and strong coupling problems.
}

\section{Introduction}
 
One of the very interesting results of astrophysical observations is that electromagnetic phenomena are present in all parts of the Universe that we can currently observe. 
While electric fields are in general suppressed due to the high conductivity of the early Universe, the existence of magnetic fields has been confirmed on almost all scales,
from planets to the intergalactic medium \cite{Grasso:2000wj, Giovannini:2003yn}.
The presence of magnetic fields in the Universe is an indication of their astrophysical and cosmological importance, which is further emphasized by the 
fact that they regulate several important processes -- such as structure formation and particle propagation. This naturally brings the question of which processes could produce magnetic fields of the 
observed strengths and correlation lengths. While various astrophysical models were developed to address this question \cite{Subramanian:2008tt, Kandus:2010nw}, the potential 
existence of magnetic fields in the voids of the intergalactic medium seems to suggest that magnetic fields could be produced cosmologically.
Lower bounds on the strength of extragalactic magnetic fields have been obtained from the non-observation of $\gamma$-ray emission  from distant TeV blazars that were expected to 
produce electromagnetic cascades in the GeV range \cite{2010Sci...328...73N}. The obtained lower bounds on extragalactic magnetic fields suggest that these fields are remnant from 
the early Universe since it is unlikely that astrophysically originated fields at the galactic scale could have an impact on the voids. However, while observations have been increasingly 
ruling out some models, it has not been possible to decisively ascertain the magnetogenesis mechanism preferred by our Universe yet \cite{Durrer:2013pga}. \\ \hfill 

In order for magnetic fields to be produced, some turbulence inducing and out-of-equilibrium process, such as a first order phase transition, was usually considered. After indications, based on the determination of the Higgs boson mass, that in the standard model quantum chromodynamics (QCD) and electroweak transitions, instead of being first order phase transitions, are crossovers, the most natural choice for magnetogenesis falls over inflation \cite{Turner:1987bw}. However, it is not straightforward to obtain magnetic fields of the strength and correlation lengths of the presently observed fields. Conformal invariance has to be broken in a conformally flat universe in order for the generated field strengths to be sufficiently high \cite{Dolgov:1993vg}. Many approaches to break the conformal invariance of the electromagnetic action during inflation are based on adding direct couplings to, for example, curvature \cite{Kunze:2009bs}, torsion \cite{doi:10.7566/JPSCP.1.013123}, scalars \cite{Ratra:1991bn, Gasperini:1995dh, Giovannini:2007rh, Giovannini:2001nh, Martin:2007ue}, pseudo-scalars \cite{Fujita:2015iga, Bamba:2014vda} and charged scalar fields \cite{Giovannini:2000dj}. However, the generated fields are very sensitive to the variation of the coupling constants such that parameters need to be fine-tuned \cite{Subramanian:2009fu}. The effects of back reaction on the background expansion from the produced scale invariant magnetic spectrum restrict the amplitude of the primordial seed magnetic fields, constraining the viability of such models \cite{Fujita:2012rb, Martin:2007ue}. 
Additionally, the need to solve the strong coupling problem at the beginning of inflationary expansion \cite{Demozzi:2009fu} usually implies more complexity, such as an extradimensional scale factor coupling \cite{Atmjeet:2014cxa}.    
\\ \hfill 

It is important to take into account the fact that cosmological magnetogenesis models are all dependent on the model assumed for cosmological evolution.
In this manner, the large majority of magnetogenesis models, such as quantum electrodynamics (QED), QCD and inflationary magnetogenesis, usually assume the validity of the standard $\Lambda$CDM model, where CDM stands for Cold Dark Matter. The standard cosmological
model was shown to be consistent with a large number of various observations and tests \cite{2012reco.book.....E}, but at the same time it cannot explain the basic aspects of the observed cosmological and astrophysical dynamics without introducing
new forms of matter-energy, termed dark-energy and dark-matter, for which -- after decades of very intensive research -- there is no direct empirical evidence. The most popular approach 
was to attribute dark energy to the cosmological constant, but within the standard model this leads to further difficulties, since it cannot be attributed to vacuum energy directly due to
its extremely small observed value \cite{Carroll:2000fy}. It is also well known that our current cosmological model leads to other problems, such
as the horizon problem, which requires some additional assumptions and the introduction of new processes to be solved within this framework. This further reduces the simplicity and
coherence of the theory. For instance, the most popular mechanism for the solution of the horizon problem is given by inflationary cosmology \cite{Baumann:2008bn}, but
it brings new -- mostly not well motivated -- additions to the model, such as a new scalar field and the coupling to it. Moreover, these additions also require fine-tuning
in order to obtain the proper spectrum and the amplitude of primordial density perturbations. On the other hand, the fundamental theoretical and conceptual limitation of the $\Lambda$CDM model,
and of Einstein's general relativity, on which it rests, is that a complete physical description of the Universe is not possible. Namely, assuming that the usual energy conditions for the cosmological matter
are satisfied and that spacetime has causally non-pathological features, it follows that the Universe needed to have an initial singularity \cite{1966RSPSA.294..511H, 1966RSPSA.295..490H, 1967RSPSA.300..187H}. 
Assuming that this singularity is physically real would mean that any physical description of the Universe would need to finish at that point. This would be contrary to the complete
development of physics, which aims at showing that the Universe can be successfully comprehended in terms of a logical and structured description. Therefore, an approach which is much more meaningful
consists of attributing these singularities in our theory of gravity to its incompleteness and to the non-applicability of general relativity to the regimes of high curvatures and high energies.
Indeed, while the fundamentals of general relativity -- like the equivalence principle and the principle of general covariance -- show important internal consistency and are of special significance -- both physically and philosophically --
there are no reasons why we should assume that the mathematical form of Einstein's field equations needs to be considered as complete, specially when going to the high energy limit. 
Indeed, we should also have on our mind that the standard form of Einstein's equations was actually derived as the simplest construction leading to the Newtonian weak field limit. Furthermore, in the high
energy regime, necessary to describe the early stages of the Universe, we should expect that a quantum gravity description becomes valid, establishing a synthesis between quantum 
principles and the principles of general relativity. As the problem of dark energy seems to suggest, these quantum effects could become significant 
already on energies much lower than the Planck scale. The future theory of quantum gravity should be expected to wipe out the singularities existing in standard general relativity, 
by virtue of the high-energy modifications that it will introduce. While this more complete theory is still unknown, the approach which seems the most justified to us is 
to consider the potential corrections to general relativity effectively, by investigating the possible modifications of the theory, which are 
in accord with the current empirical data and with already confirmed physical principles. 
With these considerations taken into account, it is very interesting to ask if the problems of magnetogenesis are actually somehow related to the foundational problems of the dominant $\Lambda$CDM 
paradigm, and if they could have their natural resolution in frameworks that go beyond the Big Bang cosmology. 
\\ \hfill 

Cyclic cosmology can be understood as a natural solution to the problems of the standard cosmological model. In this framework, the Universe
is eternal and undergoes an infinite number of cycles of contraction and expansion. By definition,
there is no beginning of the Universe in cyclic cosmology, and the horizon problem also does not appear since the correlation 
between spacetime points can be naturally established during the previous contraction cycle. Indeed, in the cyclic Universe, on an essential level, 
all spacetime points are correlated, and there are no horizons, since every event eventually intersects the observer's past or future 
light cone. What happens is just that the internal and random motions of physical systems make these correlations practically insignificant after 
a certain scale. The first physical proposals of cyclic cosmological models date to Tolmann and Lema\^itre \cite{Tolman:1931fei, 1933ASSB...53...51L}. These types of 
cyclic models, which were to be further developed in the following decades, for instance in the works of M. A. Markov 
\cite{Markov:1984ii}, are based on a spatially closed Universe configuration
of the FRWL (Friedmann-Robertson-Walker-Lama\^itre) geometry. Although spatially closed Universes are still not ruled out, current 
observations seem to favour a flat Universe geometry \cite{Okouma:2012dy,
Yu:2017iju}, thus these models based on a closed geometry are currently not in the focus of interest (although there were
recently also some new proposals regarding such models, see for instance \cite{Ellis:2015bag}). Perhaps, the most well-known model of cyclic cosmology is 
the Ekpyrotic model \cite{Khoury:2001wf}, which was developed as a potential alternative to inflation. However, it should be stated that 
there is no contradiction between the paradigm of the cyclic Universe and the inflationary phase -- i.e. the stage of inflationary expansion 
could take place as one of the phases of the cosmological cycle. Moreover, the Ekpyrotic model, being based on string theory, uses the typical elements of inflation, 
such as proper scalar potential, and it could as well be considered a special case of inflationary theories. Recently, a new approach 
was proposed in \cite{Pavlovic:2017umo} where it was discussed how the cyclical evolution can be realized in a gravity model 
containing higher order curvature quantum corrections with running couplings, as it is characteristic of effective 
field theories.\\ \hfill 

Bouncing cosmological solutions \cite{Battefeld:2014uga, Cai:2014bea, Brandenberger:2016vhg, Astashenok:2013kka, Nojiri:2017ncd}, representing the transition from a contracting to an expanding phase of the Universe, are
also a necessary ingredient of cyclic Universe models, but for that purpose, the bouncing phase needs moreover to be combined with the approximately standard 
cosmological evolution following it (radiation dominated, matter dominated, effective cosmological term dominated phases) and with a turnaround phase -- the transition from the expanding to the contracting phase, which happens at the end of each cycle. 
The problem of magnetogenesis could be one of the problems and limitations
of the Big Bang paradigm, and this naturally leads to the program of solving magnetogenesis beyond the Big Bang cosmology, i.e. in models that tend to overcome its limitations.
Magnetogenesis produced out of a non-minimal coupling has also been studied in the context of bouncing cosmologies \cite{Sriramkumar:2015yza, Koley:2016jdw, Qian:2016lbf}. Refs.~\cite{Koley:2016jdw, Qian:2016lbf} show how non-singular bouncing cosmological models, achieved by adding a bouncing term to the Lagrangian, can avoid the backreaction and strong coupling problems. 
The main problem of the various magnetogenesis models based on non-minimal coupling is that the main ingredient -- the scalar field -- is introduced without any proper physical motivation,
and its properties and the type of coupling are usually artificially constructed just to lead to the desired results. The same criticism can also be applied to various models which 
use a non-motivated coupling between curvature and vector potentials (which moreover breaks the gauge invariance of the theory and should be 
thus approached with suspicion).\\ \hfill 

In the magnetogenesis model that we propose in this work, magnetic fields are dominantly produced during the contraction phase 
of the previous cycle, and not during the bouncing phase. Assuming that the Universe enters the contraction phase approximately empty,
which solves various problems of a contracting Universe -- including the instabilities which otherwise generally arise -- we argue that 
the proper picture of the electromagnetic interaction during this epoch is given in terms of vacuum electrodynamics on curved spacetime
and not in terms of magnetohydrodynamics. We show that, in typical 
conditions for the beginning of a contraction phase, that is, the beginning of a new cycle, magnetic fields of sufficient strength to account for the present observational bounds
can be simply created, without any further assumptions and additions (such as new scalar fields, non-minimal couplings, etc.). 
\\  \hfill 

To analyse the question of magnetogenesis as proposed, we first define the background geometry on \S~\ref{sec:background} and list the Einstein-Maxwell equations used for the study of electrodynamics in curved spacetime on \S~\ref{sec:EMeqs}. We study magnetic field generation during the contraction phase of the Universe by developing the adequate formalism in \S~\ref{sec:ansatz} and presenting the solutions in \S~\ref{sec:plots}. We conclude by discussing our results in \S~\ref{sec:conclusão}.

\section{Geometry of the Contracting Universe} \label{sec:background}

The central assumption of our model is that the Universe undergoes endless cycles of contraction and expansion, so that the Big Bang is replaced by a transition from the contracting
phase of the previous cycle into the expanding phase of the following cycle -- the cosmological bounce. We assume that effects coming from the generalization of the standard Einstein equations lead to solutions of the field equations that enable both the early bouncing phase and the late transition from an accelerated expansion to the contraction of the Universe.
Such generalizations of Einstein's equations are understood to come from quantum gravity considerations and to be related to the fact that general relativity is not a complete theory of gravity and
needs to break in the regimes of high curvatures. In the same research framework, it is also possible to understand the still unanswered problems of dark energy and dark matter as coming
from the low-energy corrections to the equations of Einstein's general relativity \cite{Tsujikawa:2010zza, Bekenstein:2010pt, Joyce:2016vqv, Katsuragawa:2016yir, Lue:2003ky, Calmet:2017voc}.
It is important to note that in such a generalized setting the cosmological term is no longer necessarily a constant, but becomes a dynamic quantity \cite{
Borges:2005qs,  Shapiro:2009dh, Pan:2017ios, Pan:2017zoh}.
\\ \hfill 

As discussed in the introduction, various studies were focused on the problem of a cosmological bouncing phase, which could replace the initial singularity of standard general relativity,
investigating it in different theoretical frameworks. At the same time, recently there were significantly less studies of the turnaround phase and of viable complete models of a cyclic Universe (for a discussion and also
the analysis of the contracting phase, see Ref.~\cite{Pavlovic:2017umo}). In order to study the consequences of the cyclical cosmological evolution on the evolution of electromagnetic fields, we consider a general model, capable of representing
a very broad class of different theoretical frameworks and modifications of the action integral for the gravitational field. We assume that the Universe is given by the spatially flat
FRWL metric,
\begin{equation}
ds^{2}= -dt^2 + a^{2}(dx^2 + dy^2 +dz^2), 
\label{FRWL}
\end{equation}
where $a$ is the scale factor, with the matter-energy component described by an ideal fluid, leading to modified Friedmann equations of the form
\begin{equation}
H^{2}=\frac{\kappa}{3}(\rho + \rho_{eff}),
\label{modfri1}
\end{equation}
\begin{equation}
\dot{H}+ H^{2}=- \frac{\kappa}{6}(\rho + \rho_{eff} + 3(p + p_{eff})), 
\label{modfri2}
\end{equation}
with $\kappa= 8 \pi G$, and $H=a'(t)/a(t)$. Here $\rho_{eff}$ and $p_{eff}$ describe the effective contributions to the standard Friedmann equations coming from the modification of the Lagrangian for the
gravitational interaction or from the addition of new energy-matter components. They could, thus, potentially correspond to the energy density and pressure of some matter or scalar fields, but also
to terms coming from $f(R)$ gravity or from some more general scalar-tensor theories of modified gravity, loop quantum gravity or string theory, a time-dependent cosmological term, quintessence or phantom energy, etc.
For our present purposes of investigating electrodynamics in cyclic cosmology, the question of a concrete theoretical framework enabling a cyclic evolution of the Universe is not of interest here,
but only the general properties of the spacetime describing such Universe -- which then impose certain conditions on the possible functional forms of $\rho_{eff}$ and $p_{eff}$.  
This mathematical correspondence between a large class of modified gravitational theories and the effective $\rho_{eff}$ term is possible in a strict sense due to the equivalence of such theories with general relativity, supplemented with additional matter or scalar fields \cite{Sokolowski:1995dk, Kijowski:2016qbc}. However, we stress that such a formal equivalence does not change the fact that the physical content in both those approaches is completely different, since in the case of modified gravity theories there are no new physical substances, but the physical laws governing gravity are changed. Thus, $\rho_{eff}$ can in fact be completely unrelated with any existing substance, and just represent a collection of mathematical terms coming from the correction of the standard Lagrangian of general relativity.
\\ \hfill 

In general, in
order to have the initial singularity replaced by a cosmological bounce, there is a minimum of the scale factor $a_{min}$ to which corresponds the time $t_{min}$, around which the bounce takes place, and the solutions of the modified Friedmann equations \eqref{modfri1}-\eqref{modfri2} need to satisfy the conditions: $H(t_{min})=0$,
$H(t_{min}-d)<0$ and $H(t_{min}+d)>0$, where $d>0$ is an arbitrary time parameter, such that $|d - t_{min}|<|t_{max}-t_{min}|$, with $t_{max}$ being the time at which the scale factor reaches its
maximal value. Additionally requiring that  all physical quantities -- $\rho$, $\rho_{eff}$, $p$, $p_{eff}$ -- stay finite during the bouncing phase, it follows that $\dot{H}$ also
needs to stay finite, and it is then simply related to the value of the Ricci scalar during the bounce, $R_{\rm bounce}=6\dot{H}(t_{min})$. 
In order for this type of solutions to be possible, the requirements for the validity of singularity theorems need to be violated, i.e. 
the total contribution of the energy and pressure needs to violate the Strong Energy Condition (SEC), viz.  
$(\rho + \rho_{eff})+ (p + p_{eff})<0$. 
\\ \hfill 

The contraction phase of the Universe can lead to more physical difficulties, since the contraction of the Universe with a non-vanishing 
matter-energy distribution in general leads to growing anisotropy modes and to the Belinsky-Khalatnikov-Lifshitz (BKL) instability -- which destroys the homogeneous and isotropic 
spacetime background \cite{doi:10.1080/00018737000101171}. Furthermore, the contraction 
of a matter-dominated Universe would lead to the collapse of matter leading to a huge increase in temperature, which would then 
increase in every cycle \cite{Lifshitz:1963ps, PhysRevLett.22.1071}. This problem can in general be avoided considering the case in which the Universe starts the 
contraction approximately empty \cite{Markov:1984ii, Davies:1988dma, Pavlovic:2017umo} and the bounce happens before 
the critical matter and radiation density are reached. This condition can for instance be fulfilled if the late accelerated expansion phase 
is followed by the growth of an effective cosmological term such that it becomes dominant enough to break all the bounded systems and the particle 
distribution becomes further diluted by the subsequent expansion. Following this assumption of a nearly empty Universe at the beginning 
of contraction, viz. $\rho \approx0$ and $p \approx 0$, it follows that at the time of maximal scale factor, $t_{max}$, which is the time at 
which the Universe enters into the phase of contraction, the following conditions need to be satisfied: $H(t_{max})=0$ and $\rho_{eff}(t_{max}) + 3p_{eff}(t_{max})>0$. Comparing this last condition 
with the one at the bouncing phase, it is obvious that cyclic solutions are possible only if $\rho_{eff}$ and $p_{eff}$ have a dynamic
effective equation of state, $w=p_{eff}/\rho_{eff}$, such that it changes during a cycle -- from the values needed for the violation of the total SEC at bounce to $w_{eff}>-1/3$ 
at the turnaround. Under these conditions, the evolution of the Hubble parameter at the beginning of the contraction phase is $H= -\sqrt{k/3 \rho_{eff}}$, which is thus completely specified by 
the evolution of the modified term and, therefore, depends on the characteristic theory of modified gravity that can be used to support 
the transition into contraction. The details of such evolution are not of interest here since we are focused only on the general, model-independent and robust 
features of the contracting phase. Following this approach, we need to choose an archetypal test function, representing the $\rho_{eff}$ evolution, which is 
consistent with a contracting spacetime geometry -- that is, it needs to be a positive function with one zero-point at $t_{max}$ and a second zero point at 
the bounce, $t_{min}$. The most general and simple choice is in this setting given by a Taylor expansion of $\rho_{eff}$ to the second order with 
$\rho_{eff}(t)''<0$, namely: $\rho_{eff}(t)=\rho_{eff}'(t_{max})(t-t_{max}) + \rho_{eff}''(t_{max})(t-t_{max})^2/2$, 
where the condition $\rho_{eff}(t_{max})=0$ was used. Using the rescaling of the expansion coefficients 
$\rho'(t_{max}) \rightarrow \rho'(t_{max})(t_{min} - t_{max})\equiv b $ and $\rho''(t_{max}) \rightarrow \rho''(t_{max})(t_{min} - t_{max})^{2} \equiv b_{1}$
and with $b_{1}=-2b$, following from $\rho_{eff}(t_{min})=0$, we can write this function as
\begin{equation}
\rho_{eff}(t)= b \frac{t-t_{max}}{t_{min}-t_{max}} - b\left(\frac{t-t_{max}}{t_{min}-t_{max}}\right)^2. 
\label{effectiveterm}
\end{equation}
Such an archetypal function should be considered as an approximation suitable only for the earlier phases of the contraction period, and not 
for the evolution near to the bounce itself, since during the bounce it can be expected that the $\rho$ and $p$ will again become important and,
moreover, the modifications to general relativity will include more complicated terms capable of supporting a non-singular bounce. 
In the case of a nearly empty Universe at the earlier stages of the contracting phase, the Hubble parameter corresponding to \eqref{effectiveterm} has a simple and analytical solution. 
However, we will see that this assumption can be relaxed, taking $\rho \neq 0$, $p \neq 0$, and solving the Friedmann equation 
numerically, still using \eqref{effectiveterm}. It turns out that this does not introduce any significant difference in the evolution of the spacetime geometry nor in the evolution of the electromagnetic fields on it, as long as $\rho < \rho_{eff}$. 

\section{Electrodynamics in Curved Spacetime} \label{sec:EMeqs}

In the study of the electromagnetic interaction in the gravitational context, we make use of the electrodynamic equations applied to curved spacetime, which will be briefly described in the following, introducing the Maxwell-Einstein equations. The Lagrangian density for electrodynamics is (see e.g. \cite{Subramanian:2015lua})
\begin{equation} \label{eq:Lag}
\mathcal{L} = -\frac{1}{16\pi}F_{\mu \nu}F^{\mu \nu} + A_\mu J^\mu \,,
\end{equation}
where the first term represents the electromagnetic field tensor $F_{\mu \nu}=A_{\nu,\mu} - A_{\mu,\nu}$ and the second the interaction, with $A_\mu$ the electromagnetic four-potential and $J^\mu$ the four-current density. 

If we decompose the metric by choosing a time direction by considering the four-velocity $u^\mu=dx^\mu/ds$ of observers that measure the electromagnetic fields, with the property $u^\mu u_\mu=1$, and into the effective spatial metric for the observers orthogonal to the four-velocity,  $h_{\mu \nu}=g_{\mu \nu} + u_\mu u_\nu$. Thus the total metric can be written as
\begin{equation}
ds^2 = -(u_\mu dx^\mu)^2 + h_{\mu \nu} dx^\mu dx^\nu\,. 
\end{equation} 
The electromagnetic tensor can be related to the four-vector of magnetic and electric field by using the four-velocity of its observers, just by writing
\begin{equation}
B_\mu = \frac{1}{2}\epsilon_{\mu \nu \rho \lambda} u^\nu F^{\rho \lambda}
\end{equation}
and
\begin{equation}
E_\mu = F_{\mu \nu} u^\nu \,,
\end{equation}
respectively.

The equations that can be derived from the electromagnetic action $S= - \int \sqrt{-g} \mathcal{L} d^4x$ in terms of the electrical and magnetic fields include (following again Ref.~\cite{Subramanian:2015lua})
Gauss's law for magnetism
\begin{eqnarray} \label{eq:GaussB}
D_\beta B^\beta &=2\omega^\beta E_\beta  \,,
\end{eqnarray}
where $D_{\beta}$ is the spatial projection of the covariant derivative $D_{\beta}B^{\alpha}=h^{\mu}_{\beta}h^{\alpha}_{\nu}B^{\nu}_{;\mu}$ and  the vorticity vector, $\omega^\nu= -\omega_{\alpha \beta}\epsilon^{\alpha \beta \mu \nu}u_\mu/2$, is defined from the vorticity tensor $\omega_{\alpha \beta}$. The vorticity tensor is, by definition, the antisymmetric part of the spatial projection of the covariant derivative of the four-velocity, $D_{\beta}u_{\alpha}$. 
One can also derive Faraday's law, that takes the following form
\begin{equation} \label{eq:FaradayB}
h^\kappa_\alpha \dot{B}^\alpha =\left[\sigma^\kappa_\beta  + \omega^\kappa_\beta - \frac{2}{3}\Theta\delta^\kappa_\beta\right]B^\beta - \bar{\epsilon}^{\kappa \mu \nu}\dot{u}_{\mu}E_\nu - {\rm Curl}(E^\kappa) \,,
\end{equation}
where $\dot{u}_{\mu}=u^{\alpha}u_{\mu; \alpha}$, $\dot{B}^{\alpha}=u^{\sigma}B^{\alpha}_{; \sigma}$ and  $\sigma_{\alpha \beta}$ is the shear tensor -- the trace-free part of the symmetric component of $D_{\beta}u_{\alpha}$ -- while $\Theta=u^\mu_{;\mu}$ is the expansion scalar. 

From the relation between the electromagnetic field and its dual ${}^{*}F^{\mu \nu}=\epsilon^{\mu \nu \alpha \beta}F_{\alpha \beta}/2$, Gauss' law for the electric field and Ampère's law are obtained in the form of  
\begin{equation} \label{eq:Gauss}
D_\beta E^\beta = 4\pi\rho_q - 2\omega^\beta B_\beta\,, 
\end{equation}
with $\rho_q= -J^\mu u_\mu$, and  
\begin{equation} \label{eq:Faraday}
h^\kappa_\alpha \dot{E}^\alpha =\left[\sigma^\kappa_\beta  + \omega^\kappa_\beta - \frac{2}{3}\Theta\delta^\kappa_\beta\right]E^\beta + \bar{\epsilon}^{\kappa \mu \nu}\dot{u}B_\nu + {\rm Curl}(B^\kappa) - 4\pi j^\kappa \,,	
\end{equation}
where $j^\kappa = J^\mu h_\mu^\kappa$, respectively.

The application of these equations to the geometrical background described in \S~\ref{sec:background} will then permit us the investigation of the behaviour of electromagnetic fields in the cyclical Universe. The diverse phases of the Universe correspond understandably to very different conditions. This is thus directly reflected in the considerations and approximations that one makes in the approach to different phases. For example, in the early expansionary phase of the Universe, we customarily use magnetohydrodynamics as an adequate framework, while at the early contraction phase, this framework will not be the most suitable. Therefore, we next review the relevant conditions of the Universe at the most important moments for our later study of magnetogenesis.

\subsection{Magnetohydrodynamics in Expanding Phase} \label{sec:MHDexp}

In the early phase of expansion of the Universe, it is well known that temperature and densities were very high, which has the implication of making one describe the matter of the Universe as a hot plasma \cite{KeT}.
The conductivity of the Universe is a consequence of the differentiated force that an electric field induces in negative and positively charged particles in a plasma. The early Universe is a very good conductor, with a conductivity up to $\sigma \sim T$ \cite{Baym:1997gq}, where $T$ is the average temperature of the Universe. With such a high conductivity value, electric fields -- if present -- would have rapidly decayed due to Ohm's law. From electromagnetic fields that might have been generated before a phase of expansion, only its magnetic component will thus survive the hot early Universe.

Another well known consequence of the early Universe being approximated to a plasma is that the study of electromagnetic interactions in the early Universe requires the magnetohydrodynamic (MHD) approximation. Apart from assuming a continuum fluid, magnetohydrodynamics supposes the global neutrality of the plasma and that the current density and the electrical field are locally related. In ideal MHD, the displacement current can be neglected. These assumptions simplify the solution of Maxwell's equations coupled to the fluid equation, despite still not having a trivial solution in most cases. 
This is the common approach applied to the study of electromagnetism while the early Universe expands.   

\subsection{Vacuum Electrodynamics in Contracting Phase} \label{sec:EMcontr}

When the contracting phase begins, the Universe is in an empty state due to the expansion and consequent structure dissolution that took place before, which renders applicable vacuum electrodynamics in early contracting times. 
Unlike the hot plasma of the early phase of expansion of the Universe, that eventually cools down as it expands, the early phase of contraction will be characterised by vacuum-like conditions in the considered class of cyclic models. This is the case since after a long period of expansion the densities of matter and energy will be diluted to a negligible degree. The conductivity therefore can be taken as $\sigma =0$ in this case, although the cosmic medium continues being characterized by the presence of charged particles. The long period of expansion over which structures were disrupted has also separated those charges to a considerably high degree. The vacuum-like conditions of the medium can be understood to cause the separated charges to act as sources of electric field in the same mode as in a capacitor. Charge separation will consequently mean the presence of an electrical field. Considering the flat-spacetime form of the induction law, $\nabla \times \mathbf{B}= \partial_t \mathbf{E}$, we can roughly already assume that if there is a change in this electrical field, a magnetic field can be generated.

There is also the possibility that  the effective energy density grows sufficiently to reach the energy density that corresponds to the hydrogen ionization energy
\begin{equation}
\rho_{eff} =  n_{\rm H} W^{\rm H}_{ion}\,,
\end{equation}
where $n_{\rm H}$ is the Hydrogen number density and $W^{\rm H}_{ion}$ the ionization energy. In this case, the rupture of the atoms of hydrogen will cause a rather severe charge separation, leading again to the presence of electrical fields at the beginning of the contraction phase. A simple estimate shows us how the generated electrical field strength would be related to the distance $d$ of the charges, which will depend on the scale up to which expansion takes place
\begin{equation}
E = \frac{\rho_{eff}}{n_{\rm H}q d}\,, 
\end{equation}
based on the potential energy $U=\rho_{eff}/(qn_{\rm H})$ and where $q$ represents the charge. \\ \hfill

The components of the electromagnetic field will evolve according to \eqref{eq:GaussB}-\eqref{eq:Faraday}, which we apply to the FRWL metric \eqref{FRWL}.
The evolution of the electrical field in this regime will thus follow Ampère's law 
\begin{equation} \label{eq:Amp}
\partial_i E^i = 4\pi \rho_q\,,
\end{equation}
where $i$ denotes spatial coordinates, and Gauss's law
\begin{equation} \label{eq:GE}
\partial_t E^i + 3HE^i = \frac{1}{a} \epsilon^*_{ilm}\partial_l B^m - 4\pi j^i \,,
\end{equation}
where latin indexes run through the spatial coordinates and $\epsilon_{ijk}^*$ is the fully antisymmetric symbol with $\epsilon_{123}^*=1$. 
For the magnetic field, Maxwell's equations take the form 
\begin{equation} \label{eq:Far}
\partial_i B^i = 0
\end{equation}
\begin{equation} \label{eq:GB}
\partial_t B^i + 3HB^i = -\frac{1}{a}\epsilon^*_{ilm}\partial_l E^m \,.
\end{equation}

\section{Magnetic Field Generation During Contraction} \label{sec:ansatz}

In order to investigate magnetogenesis during the phase of contraction of the cyclical Universe, let us solve Maxwell's equations on the geometrical background described in \S~\ref{sec:background},
assuming vacuum electrodynamics and the considerations made in \S~\ref{sec:EMcontr}.
 
Since at this point, in order to explain the magnetogenesis problem we are interested in the time evolution of electromagnetic fields only, the spatial relation between fields is not of special interest. Since it does not affect the dynamics of the field configuration that one chooses, we thus can take a suitable spatial configuration given by 
\begin{equation} \label{eq:Ecomp}
E_i(i, l, m) = \phi(t)E_i(l,m)\,,
\end{equation} 
\begin{equation}  \label{eq:Bcomp}
B_i(i,l,m) = \psi(t)B_i(l,m)\,,
\end{equation}
where $\phi(t)$ and $\psi(t)$ functions of time, such that the time and space evolution of fields are independent. We moreover choose the spatial components of the fields to be  
\begin{equation}
u = \frac{\partial_m E(i,m) - \partial_l E(i,l)}{B(l, m)}\,, w = \frac{\partial_l B(i,l) - \partial_m E(i,m)}{B(l, m)}\,,
\end{equation}  
where $u$ and $w$ are constants, which need to be related through $u=-v$. 
Inserting \eqref{eq:Ecomp} and \eqref{eq:Bcomp} in \eqref{eq:GE} and \eqref{eq:GB}, while using this field configuration and considering the vacuum case, we obtain the following differential equations, respectively
\begin{equation} \label{eq:phi}
\dot{\phi}(t) +  3H\phi(t) = w \psi(t)\,,
\end{equation}
\begin{equation} \label{eq:psi}
\dot{\psi}(t) + 3H\psi(t) = u \phi(t)\,.
\end{equation}
This set of equations can be solved using the conditions adequate to the contraction phase and the Hubble parameter thereof. 

The Hubble parameter can be computed from Friedmann's equation \eqref{modfri1},
where the energy density can be modelled as $\rho = \rho_{mat} + \rho_{rad} = \rho_{mat}^0/a^3 + \rho_{rad}^0/a^4$, where $\rho_{mat}^0$ and $\rho_{rad}^0$ are the values of the matter and radiation densities at the beginning of the contraction phase, and the pressure $p = \rho_{rad}/3$. As discussed previously, by the end of contraction all structures of the Universe will have been ripped apart and matter diluted to such extent that pressure and density can be effectively taken as vanishing. If $\rho = p = 0$, \eqref{modfri1} has an analytical solution. For completion, we considered also initial matter and radiation densities and solved \eqref{modfri1} numerically, finding that, for small values of $\rho_{mat}^0$ and $\rho_{rad}^0$ consistent with the assumption of a nearly empty Universe at the beginning of the contraction phase, 
the analytical approximation provides a very good description. 

As mentioned in \S~\ref{sec:background}, at the beginning of the contraction phase, the Universe turns from expanding into contracting, implying a maximum of the scale factor at a certain time $t_{max}$. At the end of contraction, when the Universe bounces back into expansion, the scale factor will display a minimum, at a time $t_{min}$. These mean that the Hubble parameter will vanish at the points of beginning and end of the contracting phase and should be negative in between. We use \eqref{effectiveterm} in \eqref{modfri1}, encapsulating the modification of gravity that our model allows for through the effective energy density $\rho_{eff}$ that represents the effect of the  non-Hilbert component in the gravitational action.
The Ricci scalar can be easily obtained from it as well as the scale factor, since $R= 6\dot{H} + 12 H^2$ and $H \equiv \dot{a}(t)/a(t)$, respectively. 

To use dimensionless quantities, we set  $H \rightarrow H t_{min}$, $R  \rightarrow R t_{min}^2$, $u \rightarrow ut_{min}$ and normalize the scale factor to its minimum $a/a_{min}$.

\subsection{Electromagnetic Field Evolution} \label{sec:plots}

We will choose to study the evolution of the electromagnetic field by solving its time-dependent parts. 
The following initial conditions were used: the seed electrical field present through the mechanism described in \S~\ref{sec:EMcontr}, has a hypothetical strength of $E(t_{max}) = 10^{-4}$ V/m; the initial magnetic field is vanishing, $B(t_{max})=0$; and $u=1$. We have furthermore to take a value for $b$, which is chosen through the following reasoning.\\ 

The coupled equations \eqref{eq:phi} and \eqref{eq:psi} are a stable system with respect to initial conditions, as we will see, and we have freedom in setting the parameter $b$, which is directly related to the Hubble parameter evolution. This parameter has no influence in the shape of the curves $\phi(t)$ and $\psi(t)$, but it determines the strength of the generated magnetic field. Using the simplified assumption that magnetic fields only decayed with expansion, one can infer roughly the magnetic field strengths that were necessary to be present or generated at the early Universe in order for the current limits placed on extragalactic magnetic fields to be obtained.  
There have been no direct observations of extragalactic magnetic fields, but several methods allow us to set limits on them. The lower bounds alow to infer from TeV blazars emission that presently $B(t_0) \gtrsim 10^{-12}$ T \cite{2010Sci...328...73N}. Upper bounds depend on the coherence length of the field, but, at the limit, their highest value cannot exceed the observations of Zeeman splitting from the radiation of distant quasars that points to a magnetic field $B(t_0) \lesssim 0.1$ T \cite{Wolfe:2008nk}.  Since $B^\mu B_\mu \propto a(t)^{-4}$ and estimating that the present scale factor is $10^{32}$ times larger than at the Planck scale, if the mentioned bounds are taken, then  $10^{55}\gtrsim B(t_{Pl})\gtrsim 10^{44}$ T, where $t_{Pl}$ represents the Planck time. 
If we assume that the end of the contraction phase corresponds roughly to the smallest scale we know of, viz. the Planck scale, we have an estimate of the field strengths expected that this magnetogenesis mechanism supplies in order for it to be a viable hypothesis.
Let us consider the parameter 
$b'_{\rm up/low} = \kappa^{-1} b_{\rm up/low}$ s$^{-3}$ of \eqref{effectiveterm} from which the Hubble parameter is computed. We obtain magnetic fields of the strength expected to provide the current lower and upper bounds on extragalactic magnetic fields taking $b'_{\rm low} = 2.7\times 10^{4}$ and $b'_{\rm up} = 4\times 10^{4}$, respectively. In the following plots, the different choice of upper and lower bound parameters is represented in solid and dashed curves, respectively, to illustrate these two different cases. \\ \hfill

\begin{figure}[tbp]
  \centering
\includegraphics[width=0.45\textwidth]{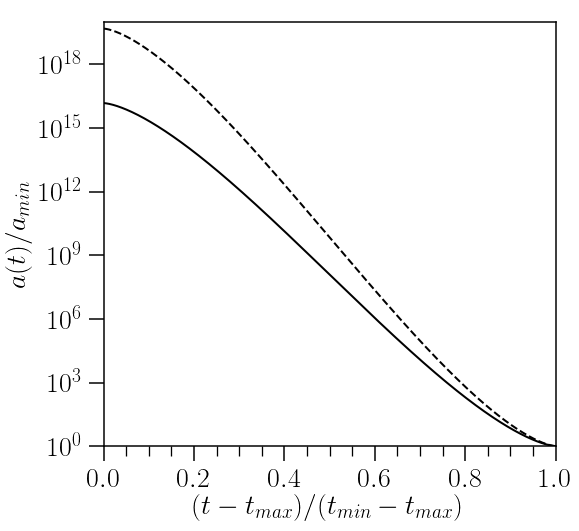} 
\includegraphics[width=0.45\textwidth]{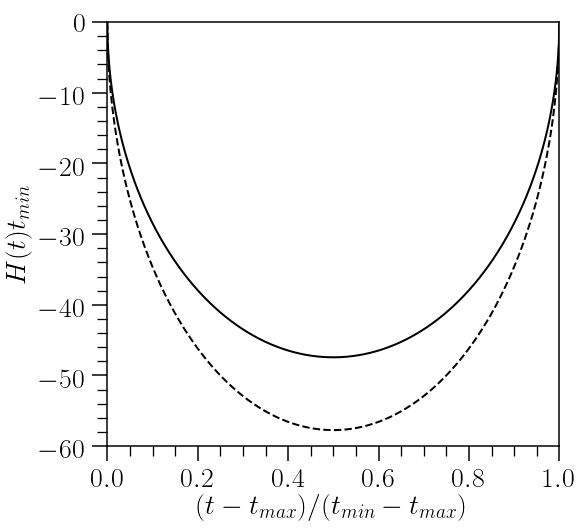} 
  \caption{\label{pic:scalefactor} Scale factor and Hubble parameter during the contraction phase of the Universe. Computed from \eqref{modfri1} using \eqref{effectiveterm}, taking $u=-w=1$. Solid lines represent the possible parameter choice in \eqref{effectiveterm} rendering the lowest possible bound on extragalactic magnetic fields and dashed lines the highest, illustrating the range of normalized values of the scale factor and Hubble parameter in these cases.}
\end{figure}

\begin{figure}[tbp]
  \centering
{\includegraphics[width=0.45\textwidth]{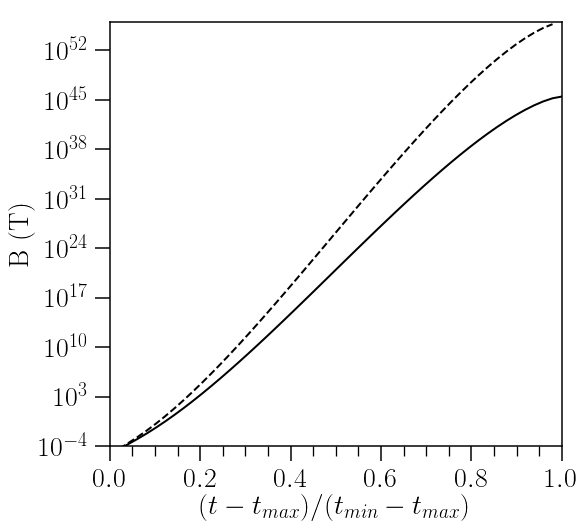} 
\includegraphics[width=0.45\textwidth]{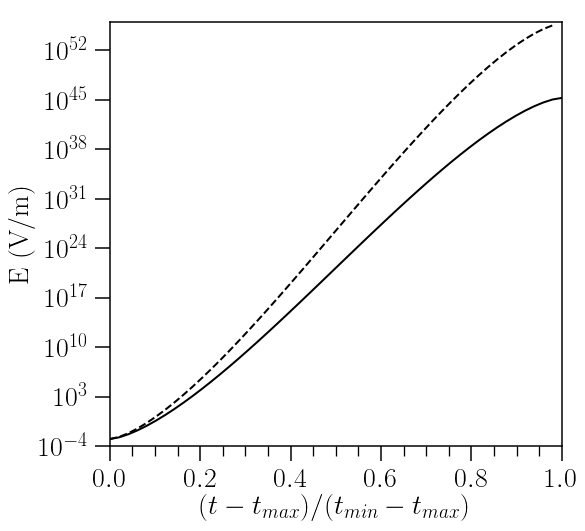}}  
\caption{\label{pic:fields} Magnetic and electrical field evolution during the contraction phase of the Universe. Computed solving \eqref{eq:psi} and \eqref{eq:phi}, respectively, taking $u=-w=1$ and using as initial conditions $B(t_{max})=0$ T and $E(t_{max}) = 10^{-4}$ V/m, while the Hubble parameter is determined through the solution of \eqref{modfri1}. Solid lines represent the possible parameter choice in \eqref{effectiveterm}, rendering the lowest possible bound on extragalactic magnetic fields and dashed lines the highest, illustrating the allowed range of generated fields in these cases.}
\end{figure}

Figure~\ref{pic:scalefactor} shows on the right panel the modelled Hubble parameter in this case, according to \eqref{modfri1} and on the left panel the scale factor obtained by integrating $H(t)$. Both show the expected discussed behaviour for the contraction regime. \\ \hfill

As can be concluded through Figure~\ref{pic:fields}, in this setting, from a small seed electrical field a strong enough current is generated that induces a magnetic field. This field will not only be sustained but will also get amplified in the course of contraction due to the decrease of the scale factor thereof. \\ \hfill

The starting assumption used in our analysis was that the contracting Universe, which is nearly empty and permeated with weak electric fields, can be described by the FRWL metric \eqref{FRWL}. A Universe consisting only of electromagnetic fields is not consistent with the symmetry of FRWL metric and can thus not be described by the Friedmann equations. Therefore, although approximately empty, the matter-energy content of the Universe still needs to dominate over the contribution coming from the electromagnetic fields, while being significantly smaller than the contribution of the effective energy density, modelling the effects of modified gravity. If this is the case at the beginning of contraction, it is simple to see that it will remain to be so. In fact, this will happen as long as the initial density values are $\rho_{mat, rad}^0\lesssim 10^{-3}b$. Using the characteristic scalings for energy densities, at later stages of contraction $[\rho_{B}(t)+\rho_{E}(t)]/[\rho_{rad}(t)+ \rho_{mat}(t)] \approx [\rho_{B}(t_{max})+\rho_{E}(t_{max})]/\rho_{rad}(t_{max})$. Therefore, under this assumption for initial conditions, $[\rho_{B}(t_{max})+\rho_{E}(t_{max})]/\rho_{rad}(t_{max}) \ll 1$, there is no backreaction problem in the proposed model.

\section{Conclusions} \label{sec:conclusão}

Let us state the assumptions at the base of this work in order to understand the applicability scope and limitations of the same. We have assumed that electromagnetic fields are independent of the spacetime background, not affecting the metric. Isotropic and homogeneous conditions are taken and thus an FRWL metric is used. \\ \hfill

In this paper we have considered a cyclical -- non-singular bouncing -- cosmological model. Unlike the standard cosmological model, this assumes that instead of an initial singularity at the beginning of spacetime (and of the Universe as we know it), which expanded into the current observable Universe, there was a bounce in which the Universe turned from a phase of contraction into expansion and that at a certain time, the expansion will turnaround and lead into another contraction phase of the cycle. The motivation for modifying the gravitational theory is based on the incompleteness that remains when building a cosmological model based on it. Among the study of several features that accrue from cyclically viable cosmological models, we focus on the possibility of generating primordial magnetic fields -- whose creation is not a clear and understood problem -- before the early Universe. In fact, we study the conditions that would be present at the beginning of a contraction phase and deduce that the generation of electromagnetic fields would naturally occur then. As explained, a viable cyclical model requires that the contraction phase occurs after a period of sufficient expansion such that structures would have been disrupted and the Universe energy and matter densities are negligible. These conditions, apart from inducing initial electrical fields, are also favourable to the growth of magnetic fields in the subsequent contraction. We did not assume any particular modified gravity framework nor model and kept the analysis general enough to suit a broad class of hypotheses. \\ \hfill

Comparing with the existing literature, our magnetogenesis model assumes no scalar or other field, as well as no additional couplings, to trigger magnetic field production. This model also has the advantage of not causing the backreaction of the fields. We describe a cyclical Universe that will behave so by the introduction of an effective energy density that will contribute to the Hubble parameter, which can potentially account for the introduction of quantum effects into gravitation. We solve the relevant Maxwell's equations in curved space in a simplified way that allows us to obtain the evolution of electromagnetic fields in time.  Through the analysis of these solutions, it gets clear that the contraction phase of the Universe is a privileged environment for magnetic field growth. \\ \hfill
 
The main new conclusion of this work can be summarized by stating that in the cyclical Universe, magnetogenesis can occur during the phase of contraction resorting only to a small seed electrical field, which is naturally created in the conditions characterizing the beginning of the contraction phase.
The magnetic field created during contraction undergoes large amplification depending on particular details of the geometrical evolution of the Universe. During the bounce, the magnetic field survives and is carried into the expansion phase. Identifying it with the standard cosmological early Universe, the evolution that the generated magnetic field subsequently undergoes has been previously studied. Referring to the magnetic field decay caused by expansion, we can conclude that the scope of strengths of the current observational bounds on magnetic fields are totally well accounted for invoking the simple mechanism that we propose.

\section*{Acknowledgments}
This work was partially supported by the Centre for Space Research of the Faculty of Natural and Agricultural Sciences of North-West University. NL wishes to thank the warm hospitality received at the Centre for Physics of the University of Coimbra and interesting discussions with Alex Blin.

\bibliographystyle{unsrt}
\bibliography{references}
\end{document}